%% file: Kienzle-ClimateEffectsOnMBBTP4Arxiv.tex
\documentclass{article}
\usepackage[dvips]{color}

\usepackage{natbib}
\usepackage{setspace}
\usepackage{amsmath}
\usepackage{graphicx}
\usepackage{etoolbox}
\makeatletter
\patchcmd{\Ginclude@eps}{"#1"}{#1}{}{}
\makeatother

\usepackage{pifont}
\usepackage{verbatim}
\usepackage{rotating}
\usepackage{multirow}
\usepackage{longtable}
\usepackage{booktabs}
\usepackage{subfigure}
\usepackage{lineno}

\begin{document}
%\linenumbers

\title{Rising temperatures increased recruitment of brown tiger prawn ({\it Penaeus esculentus}) in Moreton Bay (Australia)}

\author{Marco Kienzle\footnote{Queensland Department of Agriculture and Fisheries, Ecosciences Precinct, Joe Baker St, Dutton Park, Brisbane, QLD 4102, Australia; University of Queensland, School of Agriculture and Food Sciences, St. Lucia, QLD 4072, Australia; Corresponding author. Tel.: +61 (0)7 3255 4232; fax: +61 (0)7 3844 8235. E-mail address: Marco.Kienzle@daf.qld.gov.au} and David Sterling\footnote{Sterling Trawl Gear Services, Manly, QLD, Australia}} %, Anthony Courtney\footnote{DAFF agri-science, Brisbane, Queensland, Australia.} and Dave Sterling\footnote{Moreton Bay Seafood Industry Assocation, Brisbane, Queensland, Australia.}}

\maketitle

%\tableofcontents

\abstract{\input{abstract.tex}}
%\keywords{Monte carlo, simulation, exponential decay function, exponential probability distribution function, linear regression, non-linear regression, maximum likelihood, generalized linear model, cohort, mortality}

\clearpage
\newpage
%\pagecolor{blue}

%\textcolor{white}{test}

\section{Introduction} \input{introduction.tex}

\section{Materials and methods} 

   \subsection{Climate data} \input{BOMdata.tex}

   \subsection{Recruitment and spawning stock biomass estimates} \input{Data-SSBandRecruitmentEstimates.tex}

   \subsection{Stock--recruitment relationship} \input{SRR.tex}

\section{Results} 

\subsection{Characteristics of the climate data} \input{Results-TemperatureTrends.tex}

\subsection{Variables explaining recruitment variability} \input{Results-VarExplRecVar.tex}

\section{Discussion} \input{discussion.tex}

\section*{Acknowledgements} \input{acknowledgements.tex}

\clearpage
\newpage
%% Bibliography
\bibliographystyle{plainnat}
%\bibliography{/home/mkienzle/mystuff/Bibliography/long,/home/mkienzle/mystuff/Bibliography/Biblio}
\bibliography{long,Biblio}

%hello \cite{aaa}
%% \bibliography{/home/mkienzle/Bibliography/a-JLong}
%% \bibliography{/home/mkienzle/Bibliography/Biblio}
%\bibliographystyle{plain}
%\bibliographystyle{plainnat}
%\bibliographystyle{abbrv}
%\bibliographystyle{apalike}

\clearpage
\newpage
\section*{Figures}

%% %%%%% CPUE and DSITIA sampling station 
%%    \begin{figure}[h!]
%%      \begin{center}
%%        \includegraphics[scale=0.6, angle = 0]{Graphics/MarcoStations.ps}
%%        \caption{Location of the sampling site in Moreton Bay.}
%%        \label{fig:MapsOfSamplingSite}
%%      \end{center}
%%   \end{figure}

%%%%% Map of the distribution of tiger prawn around Australia
   \begin{figure}[h!]
     \begin{center}
 \includegraphics[scale=0.5, angle = 0]{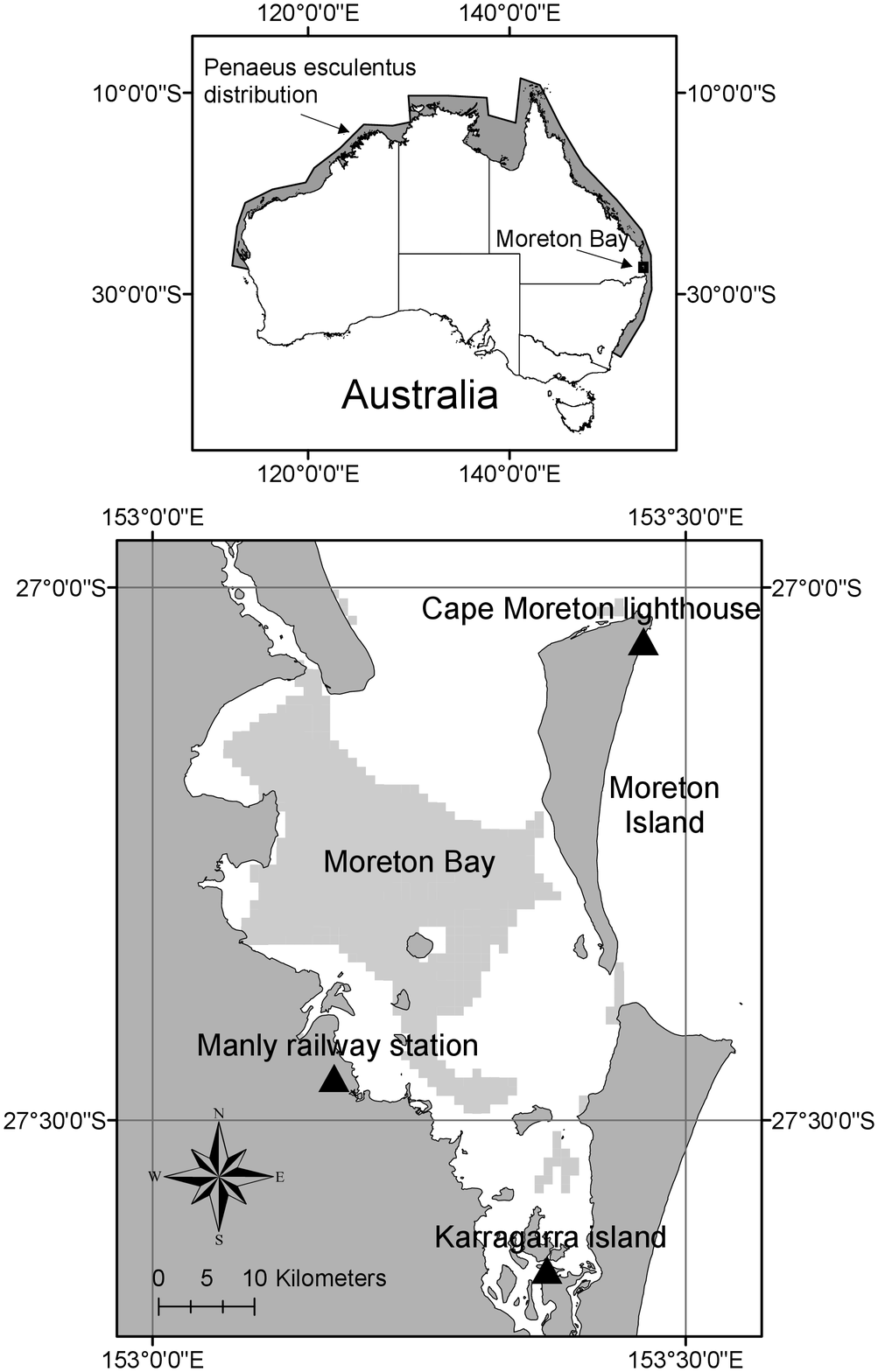}
       \caption{Top map: the spatial distribution of the brown tiger prawn ({\it Penaeus esculentus}) in Australia (from \cite{Grey83r}) with the location of Moreton Bay indicated by a black square. Bottom map: the location of trawling ground in Moreton Bay (greyed area) covering an area of about 800 km$^{2}$. Triangles indicate the location of specific Bureau of Meteorology weather stations.}
       \label{fig:Map}
     \end{center}
  \end{figure}

%% %%%%% Location of BOM weather stations
%%    \begin{figure}[h!]
%%      \begin{center}
%%        \includegraphics[scale=1, angle = 0]{Graphics/BOMstationInMoretonBay.ps}
%% %       \includegraphics[width=0.8\linewidth,bb=0 0 100 100]{Graphics/BOMstationInMoretonBay.pdf}
%%        \caption{Location of the Bureau of Meteorology weather stations.}
%%        \label{fig:LocationOfBOMWeatherStations}
%%      \end{center}
%%   \end{figure}

%%%%% Recruitment, stock recruitment estimates

  \begin{figure}[!ht]
 \subfigure[]{ % FROM THE SUBFIGURE PACKAGE
    \label{fig:RecruitmentTS}
    \begin{minipage}[b]{0.5\textwidth}
     \includegraphics[scale=0.3, angle = -90]{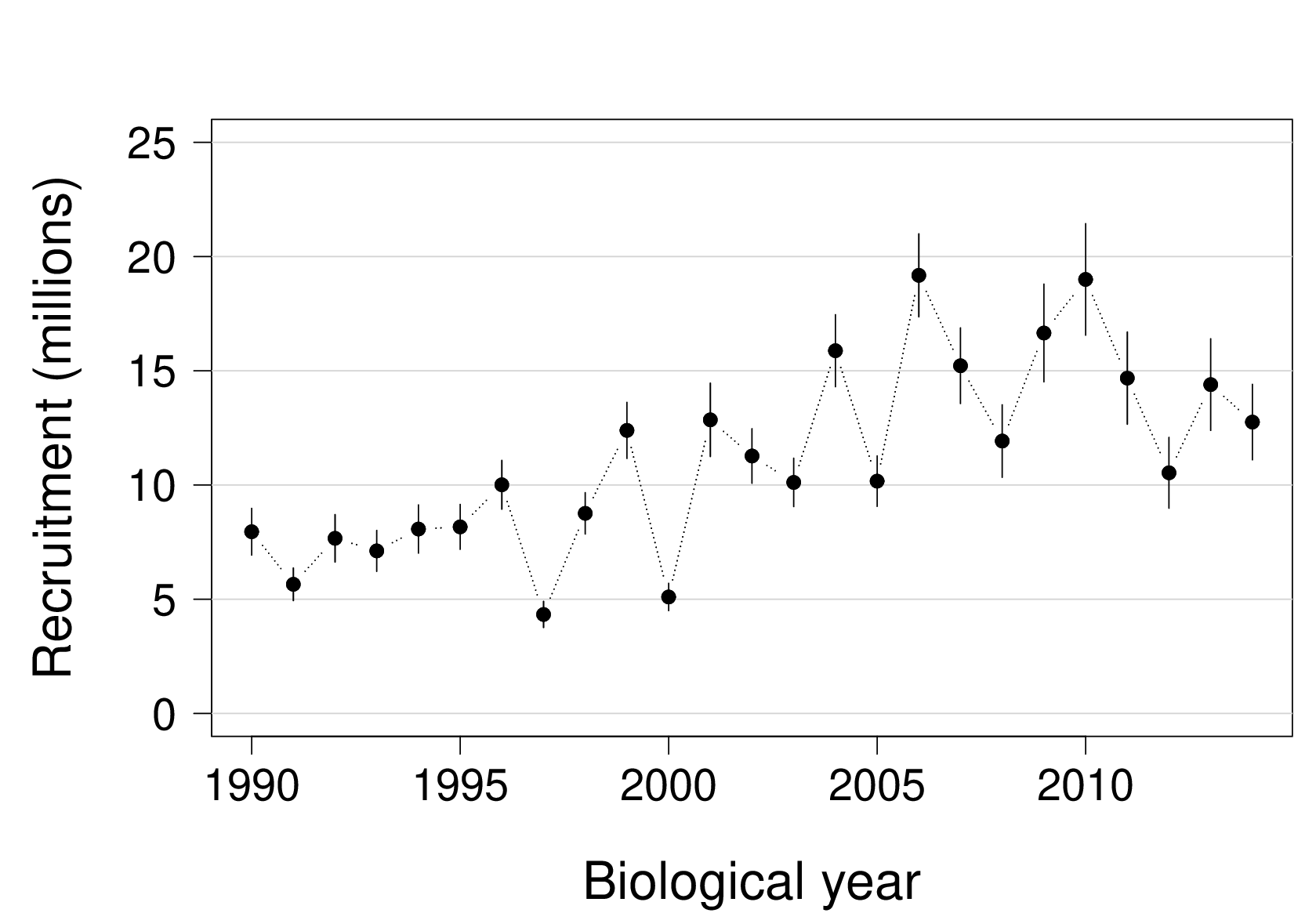}
   \end{minipage}
}
 \subfigure[]{ % FROM THE SUBFIGURE PACKAGE
    \label{fig:RickerSRR-LinearRegression}
    \begin{minipage}[b]{0.5\textwidth}
    \includegraphics[scale=0.3, angle = -90]{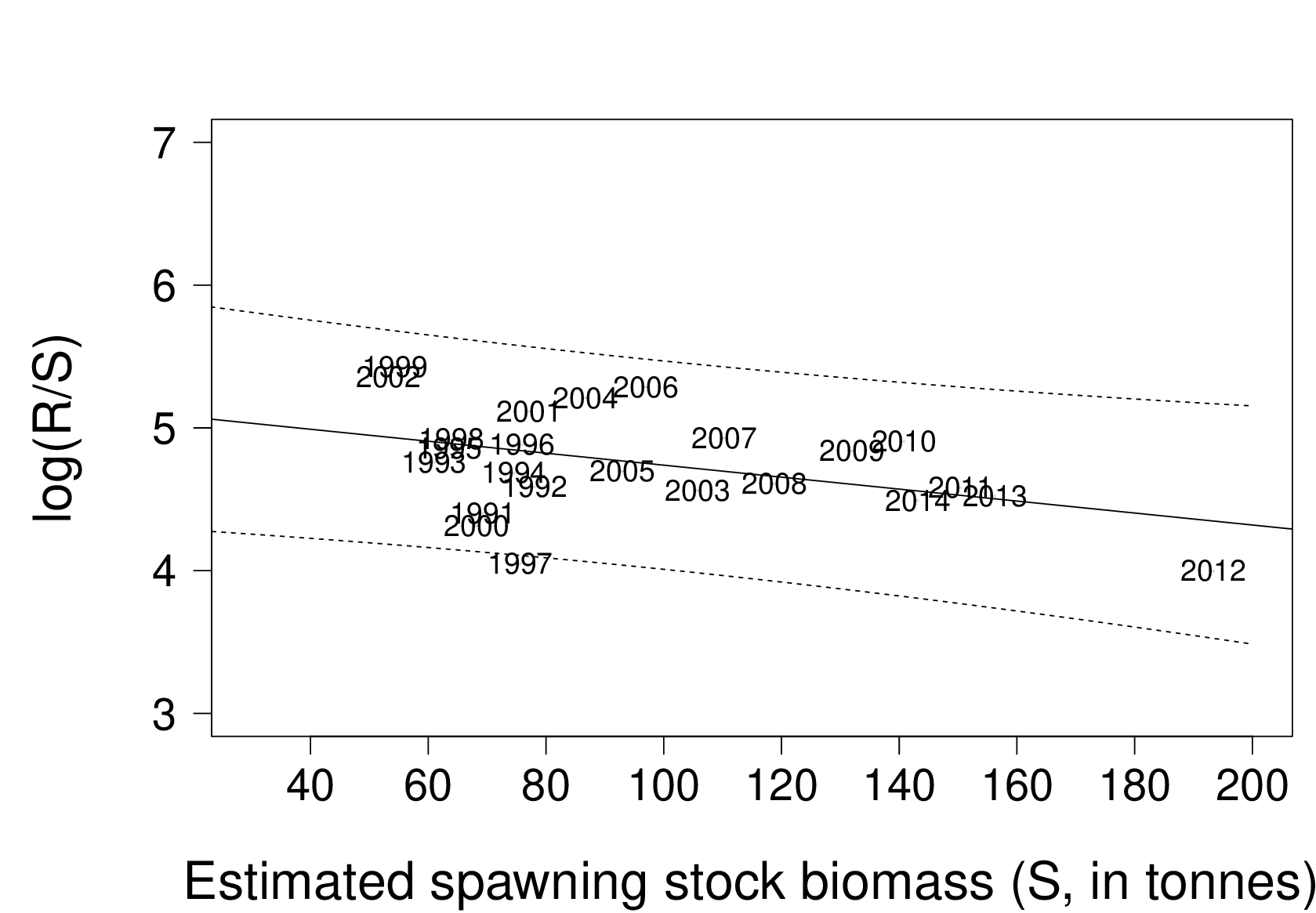}
   \end{minipage}
}
 \subfigure[]{ % FROM THE SUBFIGURE PACKAGE
    \label{fig:TS-logRecResiduals}
    \begin{minipage}[b]{0.5\textwidth}
    \includegraphics[scale=0.3, angle = -90]{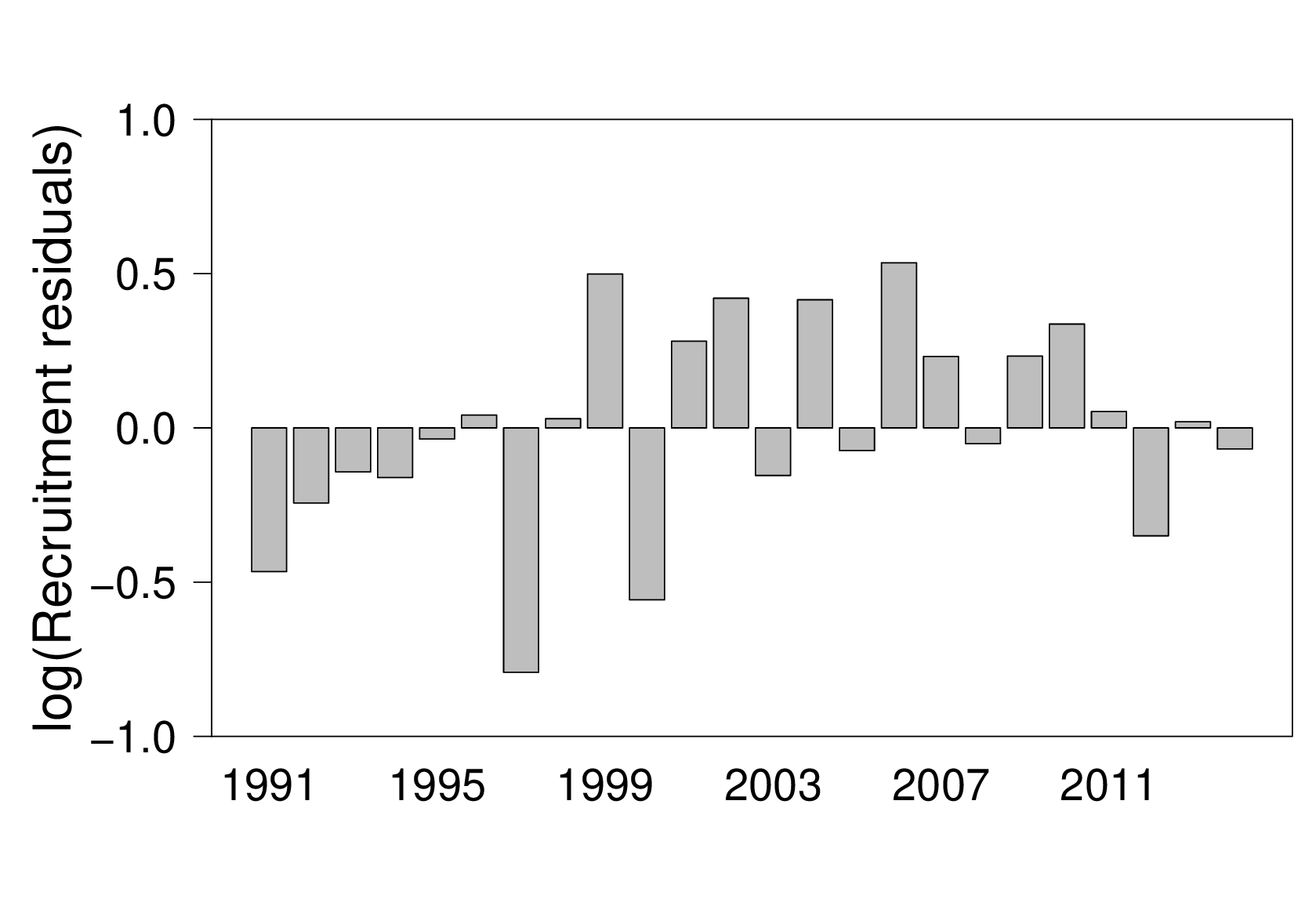}
   \end{minipage}
}
%%  \subfigure[]{ % FROM THE SUBFIGURE PACKAGE
%%     \label{fig:d}
%%     \begin{minipage}[b]{0.5\textwidth}
%%     \includegraphics[scale=0.3, angle = -90]{../Analysis/Results/HillIndexOfAvailability.ps}
%%    \end{minipage}
%% }
 \subfigure[]{ % FROM THE SUBFIGURE PACKAGE
    \label{fig:RickerSRR-SingleCurve}
    \begin{minipage}[b]{0.5\textwidth}
    \includegraphics[scale=0.3, angle = -90]{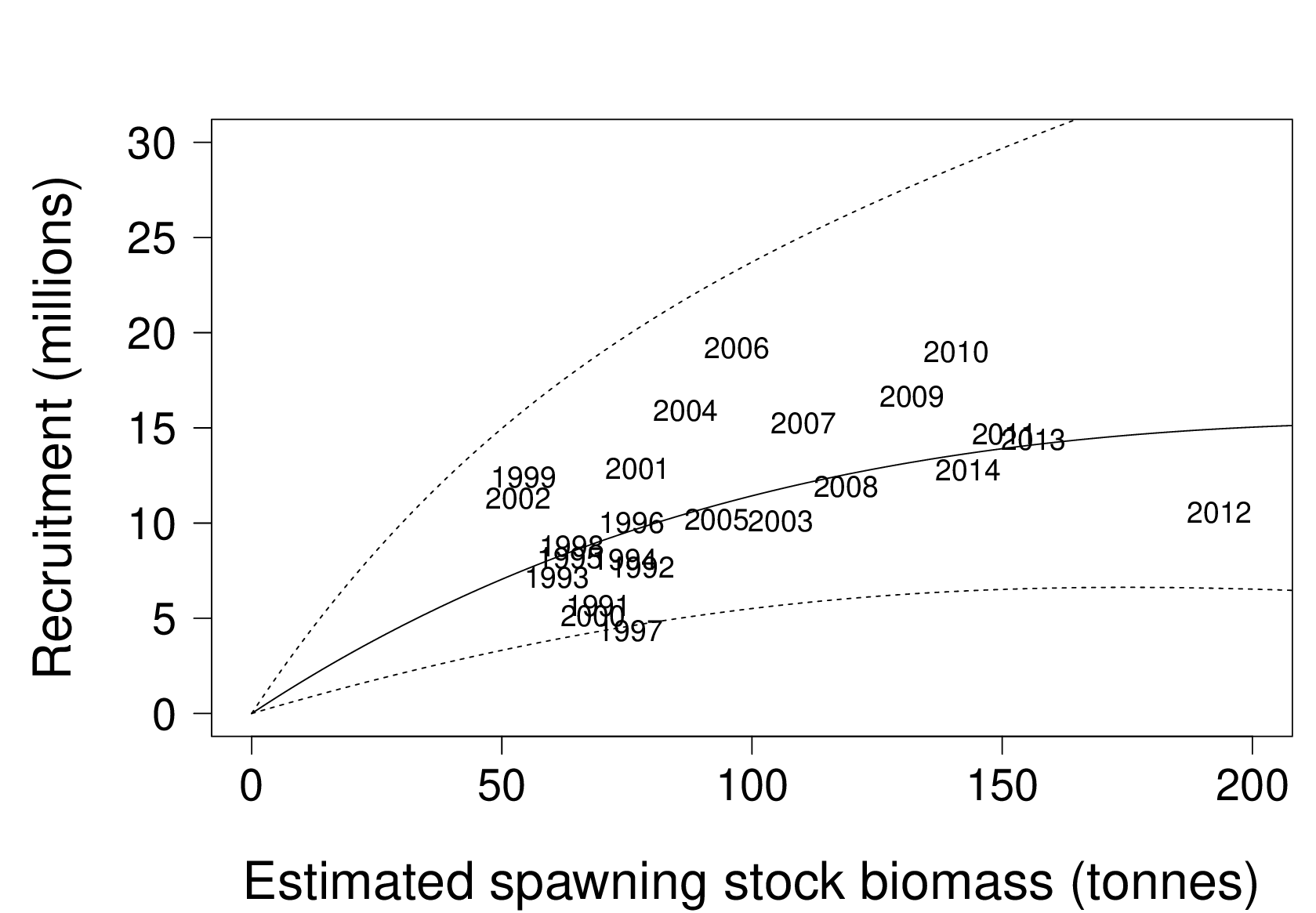}
   \end{minipage}
}
     \caption{Estimated recruitment and stock-recruitment relationship: (a) time series of brown tiger prawn recruitment in Moreton Bay ($\pm$ 1 SD) estimated by the delay difference stock assessment model \citep{KienzleEtAl2015}; (b) estimate of the stock-recruitment relationship using linear regression \citep{hil92b}; (c) residuals of the linear fit and (d) estimated relationship between recruitment ($R$) and spawning stock biomass ($S$).}

    \label{fig:EstimatesOfAvailability}
  \end{figure}

%%%%% Time series of air temperature
   \begin{figure}[!ht]
     \begin{center}
\includegraphics[scale=0.4, angle = -90]{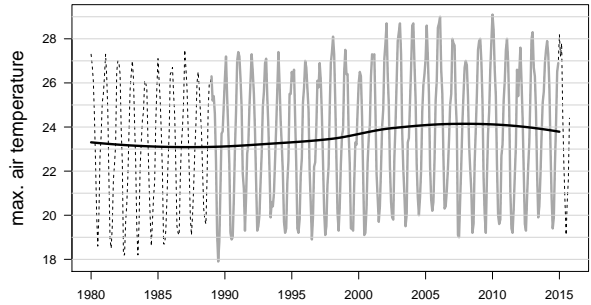}
       \caption{Time series of monthly average maximum air temperature at Cape Moreton lighthouse station from 1980 to 2015 (grey dashed line). The period overlapping the tiger prawn stock assessment are shown with a grey solid line. The smooth polynomial (black solid line) indicates the trend over the entire period.}
       \label{fig:TS-MCtemperature}
     \end{center}
  \end{figure}

%%%%% Comparison of the effect of temperature according to the linear and parabolic models
   \begin{figure}[h!]
     \begin{center}
\includegraphics[scale=0.5, angle = -90]{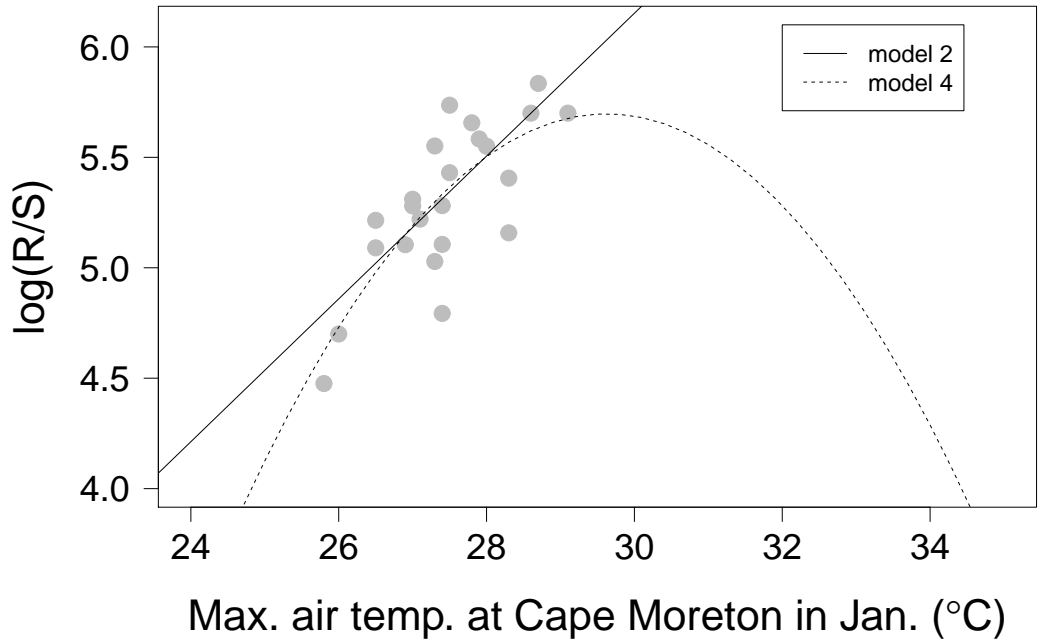}
       \caption{Comparison of the effect of temperature (x-axis) on the ratio of recruitment ($R$) to spawning stock biomass ($S$) on the log-scale (y-axis) according to the linear model (model 2) and the parabolic model (model 4).}
       \label{fig:ComparisonOfLinearAndParabolicTempEffect}
     \end{center}
  \end{figure}

%%%%% Ricker SRR estimats
  \begin{figure}[!ht]
 \subfigure[]{ % FROM THE SUBFIGURE PACKAGE
    \label{fig:LinearModelSSBandTempJanCM4logscale-BW}
    \begin{minipage}[b]{0.5\textwidth}
     \includegraphics[scale=0.4, angle = 0]{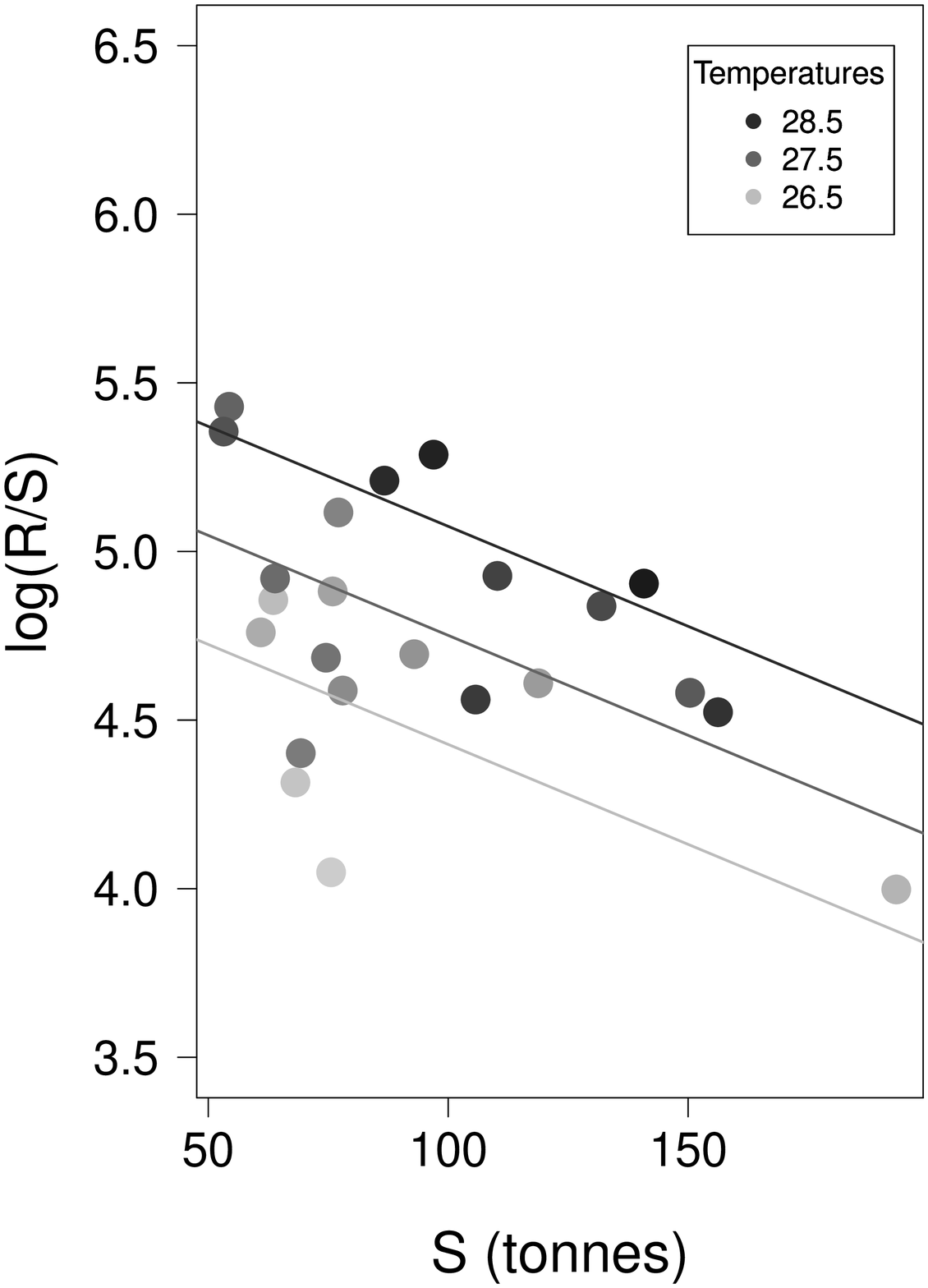}
   \end{minipage}
}
 \subfigure[]{ % FROM THE SUBFIGURE PACKAGE
    \label{fig:LinearModelSSBandTempJanCM4naturalscale-BW}
    \begin{minipage}[b]{0.5\textwidth}
    \includegraphics[scale=0.4, angle = 0]{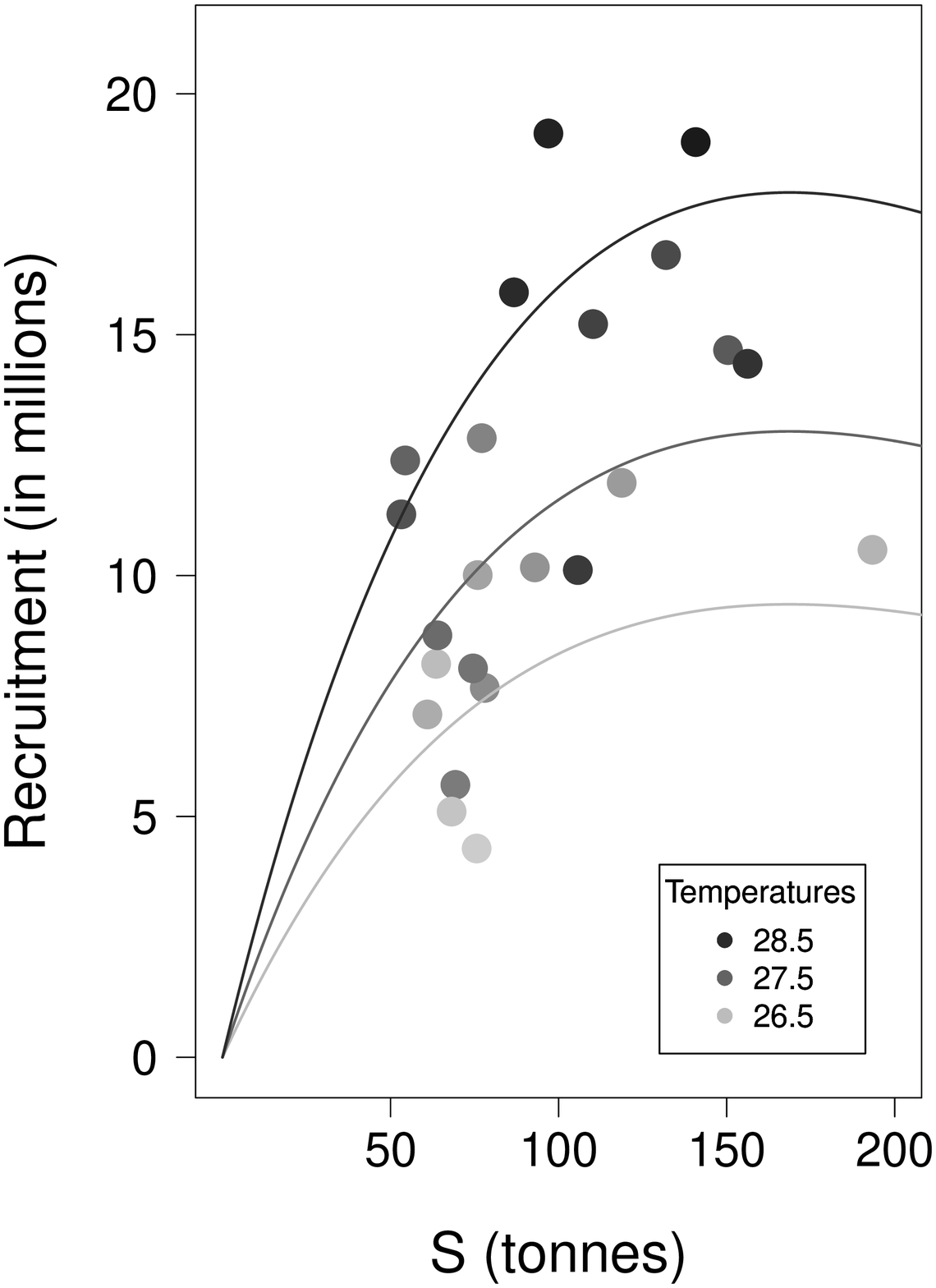}
   \end{minipage}
}
     \caption{Variation of recruitment on the log-scale (left panel) as a function of spawning stock biomass ($S$) on the x-axis and maximum air temperature measured at Cape Moreton lighthouse weather station (gray scale). Same model of recruitment on the natural scale (right panel). }

    \label{fig:EstimatesOfAvailability}
  \end{figure}

\clearpage
\newpage
%\section*{Tables}
\input{Tables/BOMstations.tex}

%\input{"/home/mkienzle/mystuff/Work/DEEDI/Biometry services/Fisheries/MoretonBay/TigerPrawn/Analysis/Results/Tables/ExampleofClimateVarCorrelation.tex"}
\input{Tables/ExampleofClimateVarCorrelation.tex}
%\input{"/home/mkienzle/mystuff/Work/DEEDI/Biometry services/Fisheries/MoretonBay/TigerPrawn/Analysis/Scripts/ClimateEffectsOnRecruitment/Results/Tables/ComparisonOfSeveralLinearModelOfRecruitment.tex"}
\input{Tables/ComparisonOfSeveralLinearModelOfRecruitmentEDITED.tex}
%\input{./Tables/DDModelPar.tex}
%% %\input{./Tables/ModelsDescription.tex}
%% \input{./Tables/HypothesesDescription4article.tex}
%% %\input{./Tables/SummaryStatistics.tex}
%% \input{./Tables/SummaryStatistics4Article.tex}

%% %\input{./Tables/ModelComparison.tex}
%% \input{./Tables/ModelComparison4Article.tex}
%% %\input{./Tables/EffectOfSwitchingHypothesis.tex}

%% %% \input{/home/kienzlem/Article-in-preparation/Mackerel-Assessment-WG/Tables/SummaryTable.tex}
%% %% %
%% %\input{}

\end{document}

%% file: abstract.tex
Abiotic factors are fundamental drivers of the dynamics of wild marine fish populations. Identifying and quantifying their influence on species targeted by the fishing industry is difficult and very important for managing fisheries in a changing climate. Using multiple regression, we investigated the influence of both temperature and rainfall on the variability of recruitment of a tropical species, the brown tiger prawn ({\it Penaeus esculentus}), in Moreton Bay which is located near the southern limit of its distribution on the east coast of Australia. A step-wise selection between 51 environmental variables identified that variations in recruitment from 1990 to 2014 were best explained by a combination of temperature and spawning stock biomass. Temperature explains 35\% of recruitment variability and spawning stock biomass 33\%. This analysis suggests that increasing temperatures have increased recruitment of brown tiger prawn in Moreton Bay.

%% file: introduction.tex
There is little doubt in the minds of fishermen and scientists alike that environmental conditions shape the productivity of marine fisheries. Similarly to terrestrial ecosystems where temperature and rainfall are key factors influencing the distribution and productivity of living organisms, marine ecosystems also change in response to climate \citep{lehodey2006a,cheung2013signature}. This is exemplified by the relationship between anchovy production along the Peruvian/Chilean coast and the El Ni\~{n}o Southern Oscillation. Identifying such ecological dependencies is expected to improve our understanding of marine ecological processes, ability to manage the exploitation of natural resources and further our capacity to adapt to climate change. %To avoid this problem, we propose in this paper to investigate the effects of various climate variables on different aspects of the dynamic of a fishery using a single-species stock assessment model to control for all important fishery mechanisms known to date. The method was applied to the brown tiger prawn ({\it Penaeus esculentus}) stock located in Moreton Bay on the east-coast of Australia.\\ 
The productivity of marine waters off Australia is generally low \citep{Haddon2007c}. A number of physical factors contribute to low concentrations of nutrients including low run-off from rivers carrying sediments from a remarkably dry continent; coastal currents flowing south from nutrient-poor northern tropical waters and lack of permanent up-welling that would recirculate nutrients. Australia is as much a desert at sea as it is on land \citep{Kailola1993}. Coastal fisheries concentrate around estuaries that provide habitats for a diverse range of species \citep{blaberb2000}. Their productivity depends on the quality and extent of habitats for new recruits \citep{Loneragan201346}; freshwater input driven by rainfall variations \citep{AEC:AEC975} and optimal temperatures \citep{Cap2014R}. \\%but, in many cases, correlations between salinity or rainfall and recruitment performed poorly when retested \citep{Myers:1998}.\\

Moreton Bay is a large estuary--fed bay in south-east Queensland, Australia (Fig.~\ref{fig:Map}). This shallow body of water has an average depth of 6.8 m and presents a gradient of salinity from brackish water in the west to oceanic in the east influenced by freshwater discharge from rivers \citep{young78a}. Sea water temperature in Moreton Bay varies throughout the year between 11.2 and 32.2 $^{o}$C: it cools/warms faster in winter/summer than the adjacent portion of ocean \citep{KienzleEtAl2015}. Moreton Bay's semi-tropical maritime climate is characterized by a wet season between October and May with an average 1500--1600 mm of rain per year \citep{nla.cat-vn5267095}. Several commercial fisheries operate in the Moreton Bay \citep{parke2013b} including a fleet of prawn trawlers catching at least five species of prawn (greasyback, {\it Metapenaeus bennettae}; eastern king, {\it Melicertus plebejus}; brown tiger, {\it Penaeus esculentus}; endeavour, {\it Metapenaeus endeavouri} and banana prawns, {\it Fenneropenaeus merguiensis}). This fishery is the southernmost prawn fishery catching brown tiger prawn on the east coast of Australia. This species of prawn is tropical and is found in Moreton Bay close to the southern limit of its geographical distribution (Fig.~\ref{fig:Map}). Cool water is probably the major physical factor limiting the production of {\it P. esculentus} in South Queensland where high production of prawns is limited to a period of about 4 months per year when the mean temperature is above 25$^{o}$C and growth rates exceed 1 mm of carapace length per week \citep{OBrien94a}. Adult tiger prawn burrow into the sediment during the day and swim near the bottom at night for a duration controlled by water temperature \citep{hil85a}, probably in response to higher metabolic rates, speeding up digestion and inducing these poikilotherms to search for food more frequently in warmer water. Catch rates are strongly correlated with water temperature \citep{Kienzle2014138, SeaCRC2012}. The annual production of prawn fisheries depends almost entirely on the annual recruitment to the fishery \citep{garcia1981life}. There are sufficient indications to show that their annual production varies from year to year in relation to large scale climate factors \citep{garcia1981life}. \cite{Penn86a} suggested that heavy rainfall events from cyclones have the potential to both increase and decrease recruit survival depending upon their timing. Sea surface temperatures during the pre-recruitment period were found to have a positive effect on the tiger prawn recruitment to the Shark Bay prawn fishery (Western Australia) but not in the Exmouth Gulf fishery, 400 km north \citep{Cap2014R}. In the Exmouth Gulf, warmer temperatures preceding the spawning season were found to have a detrimental effect on recruitment the year after.\\%\cite{Loneragan201346} reported that the size of settlement habitat for prawn postlarvae and juvenile nursery ground affected recruitment, spawning stock and landings of tiger prawn.  And \cite{SeaCRC2012} reported that temperature 60 days prior to fishing had a strong positive effect on brown tiger prawn catch rates in Moreton Bay.\\

As virtually all management advice regarding marine fisheries exploitation are based on single-species stock assessments \citep{FAF:FAF12111}, these models are a logical place to integrate environmental factors affecting production and translate this knowledge into management decisions that account for ecological processes. In practice, only a small number of stock assessments use environmental variables to predict recruitment ({\it e.g.} \cite{Ulltang01081996, Myers:1998, Lehodey2008304}) because it has proven difficult to identify reliable relationships between environmental variables and fisheries production. \cite{Robins2005a} report that many researchers use correlations or regressions of varying complexity to establish empirical relationships between environmental variables and fisheries catch or catch-rates, yet ignore the fundamental processes that govern fisheries dynamics. As a consequence, these analyses often failed to withstand the test of time because of confounding factors that were not accounted for \citep{Myers:1998, Deyle2013a}. In recent years, the stock assessment of brown tiger prawn in Moreton Bay was improved substantially by including ecological processes into the population dynamic model in the shape of a temperature-induced physiological change in catchability \citep{Kienzle2014138} and variable recruitment-timing \citep{KienzleEtAl2015}. Following these steps towards developing an ecological approach to brown tiger prawn stock assessment in Moreton Bay, recruitment was the component of the population dynamic model left with the largest amount of un-explained variability. Fitting the Ricker stock-recruitment relationship (SRR) had found no significant effect of spawning stock biomass (SSB) on recruitment variability despite large variations of SSB over its recorded history \citep{Kienzle2014138} which should have provided, according to \cite{Wal&Mar2004b}, a very good fishery dataset to study recruitment. Hence, this study investigated the effect of climate data on recruitment variability to understand which variable(s) and which weather station(s) provided data that best explained the observed variation of brown tiger prawn recruitment in Moreton Bay between 1990 and 2014.\\

%% file: BOMdata.tex
Moreton Bay covers a 1500 km$^{2}$ area between Queensland's coast and Moreton Island. This stretch of ocean is not covered by global oceanic databases. Therefore a proxy for water temperature was used given that a linear relationship exists ($R^{2}=0.81$) between daily water temperature at the bottom of Moreton Bay and maximum daily air temperature recorded at Cape Moreton lighthouse (CM) by the Bureau Of Meteorology \citep{BOM}. The slope of this relationship was estimated to equal $0.89 \pm 0.05$ \citep{KienzleEtAl2015}. The Bureau Of Meteorology (BOM) has an array of weather stations scattered across Australia that record atmospheric conditions. Of all the stations in the vicinity of Moreton Bay, only Cape Moreton lighthouse had temperature records overlapping with the fishery data analyzed (1990--2014). Rainfall data were available from several stations and those from Manly railway station (MR) and Karragarra Island (KI) were used in this analysis (Fig.~\ref{fig:Map}).\\

Total rainfall (mm) and daily maximum air temperature ($^{o}$C) averaged over each month were aggregated into variables covering longer periods of time (spring, summer, autumn, August--February or January--June) corresponding to specific periods of the biological cycles of brown tiger prawn (reproductive, growth and fishing season \citep{court97a}). These environmental variables were associated with the brown tiger prawn biological year, defined as starting around 1$^{st}$ July and finishing 12 months later around 30$^{th}$ June. Biological years were numbered according to the year that included the main fishing season (Feb. -- May). A total of 51 variables representing rainfall and temperature measurements at three different weather stations were investigated (Tab.~\ref{Tab:BOMstations}).\\% Maximum daily air temperature measured at Cape Moreton is linearly related with bottom water temperature in Moreton Bay (R$^{2}=0.81$). The slope of this relationship was estimated to equal $0.89 \pm 0.05$ \citep{KienzleEtAl2015}.\\

%% file: Data-SSBandRecruitmentEstimates.tex
Recruitment to the fishery ($R$) used in this analysis was estimated by fitting a Deriso-Schnute delay-difference model to catches of brown tiger prawn in Moreton Bay between 1990 and 2014 \citep{Kienzle2014138}. This delay-difference model assumed constant growth, constant natural mortality, knife-edge gear selectivity and targeted/non-targeted catchabilities. Seasonal variations of catchabilities induced by temperature through an eco-physiological response of tiger prawn were accounted for in the stock assessment model using a sigmoid function \citep{KienzleEtAl2015} to represent variations of the duration of emergence of brown tiger prawn at different temperatures \citep{hil85a}. The stock assessment model that best fitted brown tiger prawn catches in Moreton Bay between 1990 and 2014 was used to estimate the magnitude of yearly recruitment to the fishery, providing 25 estimates of recruitment (Fig.~\ref{fig:RecruitmentTS}). %\cite{KienzleEtAl2015} showed that brown tiger prawn recruitment to this fishery had increased between 1990 and 2014 (Fig.~\ref{fig:RecruitmentTS}). \\ %This model included fishing effects in the form of weekly variation of effort; yearly variation of fishing power; targeted and non-targeted catchabilities; timing of recruitment varying from year to year; as well as environmental effects in the shape of temperature driven catchability. 

Spawning stock biomasses ($S$) were calculated using the delay difference model's estimates of biomass by applying the weekly proportion of mature females obtained from scientific surveys \citep{court97a}, assuming an even sex-ratio.

%% file: SRR.tex
The Ricker stock-recruitment function

\begin{equation}
R=a \ S \ e^{-bS}
\end{equation}

\noindent was used to represent the relationship between spawning stock biomass ($S$) in a given year and recruitment ($R$) in the following year. The steepness of the Ricker function at the origin is given by the derivative of the function evaluated at $S=0$

\begin{equation}
\frac{dR}{dS} \Bigr|_{\substack{S=0}}=a
\end{equation}

The steepness ($a$) measures the productivity of the stock. When estimated for two different environmental conditions, its ratio ($\frac{a_{1}}{a_{2}}$) quantifies the improvement/decline in productivity associated with the difference in environmental conditions, all other things being equal.\\

Parameters $a$ and $b$ were estimated using linear regression on log-transformed data \citep{hil92b}

\begin{equation}
{\rm log} (R) = {\rm log}(a) - b S + {\rm log} (S)
\end{equation}

This model was fitted with the linear regression function in R \citep{R}, using ${\rm log} (S)$ as an offset. Environmental variables were added successively to the simplest Ricker SRR model to assess the fit of all possible multiple regressions. The best multiple regression model was identified by step-wise selection, both forward and backward, using Akaike's Information Criterion (AIC) implemented in the function {\it stepAIC} available from the MASS package \citep{ven03b}.

%% file: Results-TemperatureTrends.tex
Average maximum air temperatures in the region surrounding Moreton Bay have been trending upward over the last 35 years (Fig.~\ref{fig:TS-MCtemperature}) peaking in 2006 and 2010 and slightly declining thereafter. Air temperature increased, in average, 0.180 $\pm$ 0.014 $^{o}$C per decade between 1980 and 2015. A greater increase in air temperature was observed during the period overlapping with the tiger prawn stock assessment data (1990--2014), with an average difference of 1$^{o}$C in air temperature measured at Cape Moreton between 1985--1995 and 2005--2015. \\

Measurements of climate variables are correlated in time and space precluding their combinations into multiple linear regressions prior to addressing collinearities. Temperature measurements at Cape Moreton lighthouse were highly correlated ($\rho \geq 0.7$) between successive months in winter, spring and summer (Tab.~\ref{tab:ExampleofClimateVarCorrelation}). Moreover high correlations also existed between multiple month's periods and the shorter periods they included, meaning that high temperatures in January were indicative of high temperatures throughout summer as well as throughout the pre-recruitment growth period (August to February, Tab.~\ref{tab:ExampleofClimateVarCorrelation}). Monthly rainfall at Manly railway station and Karragarra Island was also strongly correlated, indicating that weather systems influencing this region operate at a larger scale. \\ %Maximum air temperature at Cape Moreton lighthouse (CM) were negatively correlated to rainfall during the wet season because persistent cloud coverage associated with abundant rainfall prevent sun radiations reaching the ground.\\

%% file: Results-VarExplRecVar.tex
A linear regression between recruitment ($R$) and spawning stock biomass ($S$) on the log-scale (Fig.~\ref{fig:RickerSRR-LinearRegression}) was not significant (model 1 in Tab.~\ref{tab:ComparisonOfSeveralLinearModelOfRecruitment}, F(1; 22 df)=4.139, p=0.054). The residuals of this fit showed a trend with time (Fig.~\ref{fig:TS-logRecResiduals}). Comparisons between more complicated models by step-wise selection concluded that recruitment was better explained by a multiple linear regression including both the effect of spawning stock biomass ($S$) and average daily maximum air temperature measured in January at Cape Moreton lighthouse (model 2 in Tab.~\ref{tab:ComparisonOfSeveralLinearModelOfRecruitment}). This model explained 68\% of the total variability of recruitment: approx. 35\% explained by variation in temperature and 33\% by changes in spawning stock biomass. A model including a term for interaction between temperature and spawning stock biomass (model 3) was best according to AIC but the coefficients for the interaction and effect of spawning stock biomass ($S$) were not significant suggesting ($ -3.69$e-06 $\pm \ 2.24$e-06) that temperature might change the stock recruitment relationship but there was insufficient data (n=24) to quantify it precisely.\\

The linear effect of temperature is biologically unrealistic because a dome-shaped relationship will likely better represent characteristics of living organisms, such as growth or mortality, known to vary around a maximum at an optimal temperature \citep{Gillooly21092001}. Therefore, a parabolic model of temperature was fitted to the data (model 4 in Tab.~\ref{tab:ComparisonOfSeveralLinearModelOfRecruitment}) to assess the feasibility of parametrizing a biologically more meaningful model to describe the effect of temperature on brown tiger prawn recruitment. This model has a slightly larger AIC than model 3 ($\Delta$AIC=1.25). It suggested that the current temperature range observed around Moreton Bay was limited to the ascending part of parabola (Fig.~\ref{fig:ComparisonOfLinearAndParabolicTempEffect}). Therefore past and present conditions in Moreton Bay were sub-optimal for brown tiger prawn recruitment. Given the limited range of temperature observed, the parametrization of the parabolic model is uncertain as shown by the large error associated with its parameter elevated to the power of 2 ($-6.95$e-02 $\pm$ 5.57e-02). Inclusion of an interaction term between spawning stock biomass and the polynomial of temperature measurement in January in Cape Moreton (model 5) performed worst according to AIC ($\Delta$AIC=3.55).\\

The inclusion of a second environmental factor, rainfall in May at Karragarra, to explain variability in brown tiger prawn recruitment (model 6) did not improve the model fit ($\Delta$AIC=1.92, Tab.~\ref{tab:ComparisonOfSeveralLinearModelOfRecruitment}). It is concluded that this variable does not help improve our understanding of the variation of brown tiger prawn recruitment. \\

According to the linear additive model including temperature and spawning stock biomass (model 2), temperature did not modify the density-dependence effect. It increased the recruitment to the fishery (Fig.~\ref{fig:EstimatesOfAvailability}) as measured by the steepness of the Ricker stock-recruitment relationship ($a$). An increase of temperature by 1$^{o}$C was estimated to increase the recruitment to the fishery ($a$) by a factor 1.4 ($\rm{exp}(3.45$e$-01)$). The average maximum daily temperature measured in January between 1990 and 2014 ranged from 25.8 to 29.1 $^{o}$C. According to model 2, such difference in temperature explained a variation in brown tiger prawn recruitment by a factor of 3 ($\rm{exp}(3.45$e$-01 * (29.1 - 25.8))$) between the worst and the best environmental conditions prevailing during this period of time.
According to the parabolic model of temperature (model 4), maximum recruitment would be reached at 29.9 $^{o}$C (fig.~\ref{fig:ComparisonOfLinearAndParabolicTempEffect}). This model suggests recruitment would increase by a multiplicative factor of 1.5 per $^{o}$C at 27$^{o}$C down to a multiplicative factor of 1 at 29.9$^{o}$C beyond which it would decline. In other words, warming waters would increase prawn recruitment in Moreton Bay only for a limited amount of time before declining. \\ %environmental conditions worsen.  \\%they will become a viable alternative if/when a wider range of temperature variation together with their effect on recruitment will become available.\\

%The Ricker model fitted well the data (Fig.~\ref{fig:RickerSRR-LinearRegression}). It provided a relationship to link successive generations of tiger prawn in the model and perform projections of the stock under variable fishing intensities to estimate Maximum Sustainable Yield (MSY, \citep{Kienzle2014138}).

%A substantial amount of variability was not explained by the variation of SSB (how much ?).

%Recruitment residuals on the log-scale have been increasing throughout the time series (Fig.~\ref{fig:TS-logRecResiduals}).
%[ include a comparison of model with SSB alone and SSB + temperature and SSB * temperature ] 

%Including temperature in the stock-recruitment relationship permitted the detection of a significant compensatory population effect (e.g., a dome-shaped stock-recruitment function) that was masked by temperature-driven variability  \citep{Hare2010a}.

%% file: discussion.tex
% Summarize the results, are there any comparable study ?
The results of the present work suggest that environmental factors have contributed to increase the recruitment of brown tiger prawn in Moreton Bay between 1990 and 2014. The positive relationship between temperature and recruitment is consistent with the biology of this tropical species living in an area close to the southern limit of its distribution in the southern hemisphere \citep{SeaCRC2012}. \cite{cheung2013signature} proposed to use the mean temperature of the catch to study the effect of climate change on fisheries worldwide. Their results describe the tropicalisation of sub-tropical waters around the world as a result of global warming which is supported by our results. In the present analysis, temperature variations explained 35\% of recruitment variability making this variable the largest contributor to recruitment variation between 1990 and 2014. On the other hand, spawning stock biomass in a given year explained another 33\% of the variation of the number of recruits in the following year. The increasing trend in brown tiger prawn recruitment over the last 20 years was previously regarded as the consequence of declining fishing pressure \citep{Kienzle2014138}. Since the magnitude of spawning stock biomass depends on fishing mortality as well as recruitment in the same year, itself influenced by temperature, the contribution of the reduction of fishing mortality to the recovery of the stock is likely to have been less than the contribution of environmental factors. Nevertheless, this work suggests that both factors worked together to produce a higher stock size in 2014 than in 1990. The synchronicity of changes in environmental and fishing factors around 2000 is interesting and a source of confusion as to determine which factor affected most the dynamic of this stock: this is almost certainly the result of chance because there are no evidences suggesting a causal relationship between temperature and fishing effort. Previous analyses failed to find a significant relationship between stock and recruitment by regressing Ricker's function to these data \citep{Kienzle2014138}. Including temperature in the stock-recruitment relationship (model 2) permitted the detection of a significant compensatory population effect (e.g., a dome-shaped stock-recruitment function) that was masked by temperature-driven variability, as shown by \cite{Hare2010a}. Both studies focused on species at the southern limit of their distribution. In a re-analysis of environment and recruitment relationships, \cite{Myers:1998} concluded that there is one generalization that stands out: correlations for populations at the limit of a species' geographical range have often remained statistically significant when re-examined. \cite{Myers:1998}'s results suggest that the effect of temperature reported here is likely to hold into the future and therefore be a useful knowledge for the management of the brown tiger prawn fishery in Moreton Bay. The present results are also consistent with an investigation of the effect of climate change on fisheries in Western Australia \citep{Cap2014R} which reported that fishery-independent indices of recruitment for brown tiger prawn in the Shark Bay prawn fishery were positively correlated with sea surface temperature ($\rho=0.826$).\\ %The present results are also consistent with the somewhat independent study by \cite{SeaCRC2012} reporting an increase in tiger prawn catch rates explained by higher temperature two months earlier. The problem with this empirical study based on regressions, as pointed out by \cite{Robins2005a}, is that it not assigning this effect to a specific population dynamic process such as catchability as evidenced by \cite{Kienzle2014138} or recruitment as suggested by the present study and therefore might not withstand the test of time as found by \cite{Myers:1998} because their model ignores the effect of effort on abundance. \\

% What are the consequences of this effect ?
Global warming is already occurring with significant consequences for marine and freshwater resources \citep{Lough2011a}. It is predicted to continue into the foreseeable future \citep{CsiroBom2015}, therefore these results suggest that recruitment of brown tiger prawn in Moreton Bay should continue to increase in line with an expected increase in temperature. As a result, fishermen targeting this species in Moreton Bay are likely to benefit from a rising temperature induced by climate change, at least in the short term. The optimal temperature estimated by the parabolic model used in this analysis (T$_{\rm{max}}$=29.9$^{o}$C) is close to maximal temperature recorded over the period 1990--2014, suggesting that recruitment might not increase much further than already estimated. On the other hand, aquaculture experiments estimated the optimal temperature for growth and survival of brown tiger prawn between 30 and 35$^{o}$C \citep{O'Brien1994133} suggesting that there is more scope for an increase in brown tiger prawn recruitment in Moreton Bay. This somewhat optimistic view about the consequences of climate change for the Moreton Bay prawn fishery has been obtained through the narrow lens of a single species assessment analysis. Our results should be balanced by (a) the fact that beyond optimal temperature, further temperature increase will become detrimental to brown tiger prawns and hence its fishery; (b) an evaluation of the impact of climate change on all species caught in Moreton Bay to provide an holistic view of the short to medium terms costs and benefits of climate change to the fishing industry in south east Queensland and (c) the productivity of brown tiger prawn fisheries further north may reduce as temperature increases beyond the optimum for this species, redistributing brown tiger prawn production along the east coast of Australia.\\

% What is it that increase production ? increased growth or declining survival ? there must be enough food to allow sustain larger population. How do move from this empirical study to a mechanistic model of the fishery that include the effect of temperature on recruitment ?
The delay difference model used a constant somatic growth parametrization throughout the entire time series. Therefore, a change of size at age resulting in a larger catch for a given recruitment would be incorrectly interpreted by the model as an increased estimate of recruitment. Even if the delay-difference model could be modified to incorporate prawn sizes, no size data were collected to determine whether increased temperature induced faster growth or higher larval survival leading to larger recruitment, or both. Our research suggests that warmer temperature improved brown tiger prawn recruitment in Moreton Bay, but it does not elucidate by which mechanisms this was actually achieved. Another interesting observation from this research is that past increases in recruitment have been possible only because food to sustain larger cohorts was available. There is no knowledge about how organisms preyed upon by brown tiger prawns are responding to increasing temperatures and whether they will exist in sufficiently large quantity to sustain extrapolated further increase in recruitment at higher temperatures. Food and habitat shortage are recognized as major limiting factors for the expansion of invertebrate populations in the wild \citep{Loneragan201346, chi75a}.

%% file: acknowledgements.tex
Catch and effort data were provided by the State of Queensland, Australia through the Department of Agriculture and Fisheries. Formatting of the tables was greatly simplified by \cite{stargazer}. We are grateful to Dr. A.J. Courtney and several reviewers for improving earlier versions of this manuscript.

%% file: Tables/BOMstations.tex
% CREATED on 23 Feb 2015
% MODIFIED   23 Feb 2015

%\begin{table}[ht]
\begin{sidewaystable}[ht]
\begin{tabular}{|l|cccc|cccc|}
  \toprule
Weather station & \multicolumn{4}{|c|}{Avg. max. air temperature} & \multicolumn{4}{|c|}{Rainfall} \\
                & Monthly & Seasonal & Aug--Feb & Jan--Jun & Monthly & Seasonal & Aug--Feb & Jan--Jun \\
  \midrule
Cape Moreton lighthouse (CM) & \ding{51} & \ding{51} & \ding{51} & \ding{51} &           &           &           &           \\
Manly railway station (MR)   &           &           &           &           & \ding{51} & \ding{51} & \ding{51} & \ding{51} \\
Karragarra island (KI)       &           &           &           &           & \ding{51} & \ding{51} & \ding{51} & \ding{51} \\

  \bottomrule
\end{tabular}

\caption{Type and location of the BOM weather data used in the analysis.}
\label{Tab:BOMstations}
\end{sidewaystable}
%\end{table}

%% file: Tables/ExampleofClimateVarCorrelation.tex
% latex table generated in R 3.2.0 by xtable 1.8-0 package
% Tue Dec  1 11:59:04 2015
\begin{sidewaystable}[ht]
\caption{Correlations between a subset of temperature (T) and rainfall (R) variables.} 
\centering
\begin{tabular}{lcccccccc}
  \hline
 & T.Jan.CM & T.Feb.CM & T.Aug2Feb.CM & T.Summer.CM & R.Jan.MR & R.Feb.MR & R.Jan.KI & R.Feb.KI \\ 
  \hline
T.Jan.CM & 1.00 & 0.60 & 0.83 & 0.87 & 0.39 & 0.19 & 0.04 & 0.42 \\ 
  T.Feb.CM & & 1.00 & 0.80 & 0.85 & 0.07 & -0.37 & -0.13 & -0.19 \\ 
  T.Aug2Feb.CM & & & 1.00 & 0.93 & 0.22 & 0.02 & -0.06 & 0.29 \\ 
  T.Summer.CM & & & & 1.00 & 0.23 & -0.02 & 0.01 & 0.21 \\ 
  R.Jan.MR & & & & & 1.00 & -0.00 & 0.68 & 0.26 \\ 
  R.Feb.MR & & & & & & 1.00 & -0.30 & 0.85 \\ 
  R.Jan.KI & & & & & & & 1.00 & -0.05 \\ 
  R.Feb.KI & & & & & & & & 1.00 \\ 
   \hline
\end{tabular}
\label{tab:ExampleofClimateVarCorrelation}
\end{sidewaystable}

%% file: Tables/ComparisonOfSeveralLinearModelOfRecruitmentEDITED.tex
% Table created by stargazer v.5.2 by Marek Hlavac, Harvard University. E-mail: hlavac at fas.harvard.edu
% Date and time: Mon, Dec 14, 2015 - 18:23:27
% Requires LaTeX packages: rotating 
\begin{sidewaystable}[!htbp] \centering 
  \caption{Parameter estimates from single and multiple regressions of recruitment ($R$) on the log-scale as a function of temperature in January at Cape Moreton lighthouse (Temperature), rainfall in May at Karragarra island station (Rainfall) and spanwing stock biomass (S). $\Delta$ AIC is difference in Akaike Information Criteria between a model and the best model. Parameters's standard errors are given in parentheses. Symbols used for significance levels are: $^{*}$p$<$0.1; $^{**}$p$<$0.05; $^{***}$p$<$0.01.} 
  \label{tab:ComparisonOfSeveralLinearModelOfRecruitment} 
\small
\begin{tabular}{@{\extracolsep{5pt}}lcccccc} 
\\[-1.8ex]\hline 
\hline \\[-1.8ex] 
\\[-1.8ex] & \multicolumn{6}{c}{\large{Models}} \\ 
\\[-1.8ex] & (1) & (2) & (3) & (4) & (5) & (6)\\ 
\hline \\[-1.8ex] 
 Temperature &  & 3.45e-01$^{***}$ & 6.52e-01$^{***}$ & 4.16e+00 & 3.45e+00 & 3.39e-01 $^{***}$ \\ 
  &  & (6.04e-02) & (1.95e-01) & (3.05e+00) & (1.26e+01) & (6.08e-02) \\ 
  & & & & & & \\ 
 Temperature$^{2}$ &  &  &  & $-$6.95e-02 & $-$5.22e-02 &  \\ 
  &  &  &  & (5.57e-02) & (2.30e-01) &  \\ 
  & & & & & & \\ 
 Rainfall &  &  &  &  &  & $-$4.16e-04 \\ 
  &  &  &  &  &  & (4.24e-04) \\ 
  & & & & & & \\ 
 S & $-$6.506e-06$^{**}$ & $-$8.78e-06$^{***}$ & 9.20e-05   & $-$8.55e-06$^{***}$ & 2.20e-04 & $-$8.94e-06$^{***}$ \\ 
  & (2.852e-06) & (1.87e-06) & (6.12e-05) & (1.86e-06) & (2.08e-03) & (1.88e-06) \\ 
  & & & & & & \\ 
 Temperature : S &  &  & $-$3.69e-06&  & $-$1.37e-05 &  \\ 
  &  &  & (2.24e-06 ) &  & (1.52e-04) &  \\ 
  & & & & & & \\ 
 Temperature$^{2}$ : S &  &  &  &  & 1.95e-07  &  \\ 
  &  &  &  &  & ( 2.77e-06) &  \\ 
  & & & & & & \\ 
 Constant & 5.30e+00$^{***}$ & $-$3.98e+00$^{**}$ & $-$1.24e+01$^{**}$ & $-$5.62e+01 & $-$5.00e+01 & $-$3.74e+00 $^{**}$ \\ 
  & (2.36e-01) & (1.63e+00) & (5.32e+00) & (4.19e+01) & (1.71e+02) & ( 1.65e+00) \\ 
  & & & & & & \\ 
\hline \\[-1.8ex] 
Observations & 24 & 24 & 24 & 24 & 24 & 24 \\ 
R$^{2}$ & 0.158 & 0.684 & 0.716 & 0.706 & 0.721 & 0.695 \\ 
Adjusted R$^{2}$ & 0.120 & 0.654 & 0.673 & 0.662 & 0.644 & 0.650 \\ 
\shortstack[l]{Residual Std. Error \\ \hspace{0.5cm} df } & \shortstack{0.359 \\ (22)} & \shortstack{0.230 \\ (21)} & \shortstack{0.221 \\ (20)} & \shortstack{0.227 \\ (20)} & \shortstack{0.231 \\ (18)} & \shortstack{0.230 \\ (20)} \\ 
\shortstack[l]{F Statistic \\ \hspace{0.5cm} df} & \shortstack{4.139$^{*}$ \\ (1; 22)} & \shortstack{22.697$^{***}$ \\ (2; 21)} & \shortstack{16.806$^{***}$ \\ (3; 20)} & \shortstack{15.985$^{***}$ \\ (3; 20)} & \shortstack{9.326$^{***}$ \\ (5; 18)} & \shortstack{15.219$^{***}$ \\ (3; 20)} \\ 
$\Delta$ AIC & 21.5 & 1.05 & 0 & 1.25 & 3.55 & 1.92 \\
\hline 
%\hline \\[-1.8ex] 
%\textit{Note:}  & \multicolumn{6}{r}{$^{*}$p$<$0.1; $^{**}$p$<$0.05; $^{***}$p$<$0.01} \\ 
\end{tabular} 
\end{sidewaystable} 